\documentclass[11pt, oneside]{article}   	
\usepackage{geometry}                		
\usepackage{graphicx}
\usepackage[usenames,dvipsnames]{xcolor}
\usepackage{braket}
\usepackage{amssymb}
\usepackage{amsfonts}
\usepackage{amsmath}
\usepackage{bbm} 
\usepackage{bbold}
\usepackage{siunitx}
\usepackage{lineno}
  \usepackage{young}
\begin{document}
\title{Critical behavior of SU(3) lattice gauge theory with 12 light flavors
 }
\author{Y. Meurice\\
 Department of Physics and Astronomy\\
 The University of Iowa\\
 Iowa City, IA 52242, USA }
\definecolor{burnt}{cmyk}{0.2,0.8,1,0}
\def\lt{\lambda ^t}
\def\note{note}
\def\beq{\begin{equation}}
\def\enq{\end{equation}}
\newcommand{\Tr}{\text{Tr}}

\def\gf{U(N_f)_L\bigotimes U(N_f)_R}
\def\ls{\lambda_\sigma}
\def\la{\lambda_{a0}}
\def\pdp{\phi^\dagger\phi}
\def\bpi{\boldsymbol\pi}
\def\bao{\bf a_0}
\def\etp{\eta'}
\date{\today}
\maketitle
\begin{center}
Talk presented at the APS Division of Particles and Fields Meeting (DPF 2017)\\ July 31-August 4, 2017, Fermilab
\end{center}
\centerline{\bf Abstract}
\vskip10pt
It is expected that when the number of light flavors of gauge theories is increased near or beyond some critical 
value, new and interesting behavior occurs. We discuss the qualitative properties of the RG  flows for a local $SU(3)$ theory with $N_f$ light fundamental flavors for $N_f$ near 12. 
We discuss the realization of the chiral symmetry and remind that the Wigner mode seems an unlikely alternative. We propose to use a linear sigma model to describe the light $\sigma$ masses found in recent lattice calculations for $N_f$=8 and 12. 
Progress made after the conference are reported in arXiv:1709.09264, where it is claimed that the breaking of the axial $U(1)_A$ is a key ingredient to get a small $\sigma$ mass. 
For SU(3) lattice gauge theory with 12 flavors of unimproved staggered fermions, the scaling of the 
imaginary part of the zeros of the partition function in the complex coupling plane is consistent with a first order phase transition for small values of the mass. Zech Gelzer and Diego Floor are investigating the scaling near the endpoint of the line of first order phase transition in the mass-coupling plane where a light and weakly interacting scalar is expected. 
We briefly discuss recent calculations of the second-order R\'enyi entanglement entropy and estimations of the central charge using the Tensor Renormalization Group method for the 
two-dimensional O(2) model and the possibility to extend these calculations to higher dimensions.
This is an informal summary of a talk given at the 
DPF meeting at Fermilab, July 31, 2017.

\section{Dynamical mass generation and RG flows}
Dynamical mass generation is very appealing because you can get a lot of structure from a simple input. It is very common. In the standard model,  more than 98 percent of the mass of the proton comes from quark-gluon interactions. In the case of two flavors \cite{sigma,benlee}, it can  be described by an effective theory, a Yukawa coupling $g_{\sigma NN}$ between the $\sigma$ and the Nucleons: \beq m_N\sim g_{\sigma NN}.
< \sigma >.\enq
In the standard model, we have a similar equation: \beq m_f\sim g_{BEHff}<BEH>,\enq 
with $f$ a  quark or a lepton, and  $BEH$ the Brout-Englert-Higgs boson. This suggests to ask if the BEH boson is also composite. Note that for QCD, 
\beq\Gamma_\sigma \approx m_\sigma\approx \Lambda _{QCD},\enq but for an hypothetical BEH compositeness at a new high energy scale $\Lambda_{new}$, \beq\Gamma_{BEH}<13MeV\ll m_H (125 GeV) \ll \Lambda_{new} .\enq
Recent LHC results suggest that $\Lambda_{new}>2$TeV and it is tempting to interpret the relative lightness of the BEH boson as the remnant of an approximate conformal symmetry. 
This possibility has motivated lattice studies of models which are almost conformal \cite{bz,st05,ds07}. For reviews of the motivations and the recent lattice literature, see Refs. \cite{bsr,dgrmp,nogrev}. A popular setup is a $SU(3)$ gauge group of metacolor with $N_f$  light fundamental Dirac metaquarks. 
The 2-loop perturbative beta function has a nontrivial zero for $N_f>8$ \cite{bz}. The values $\alpha_c$ at which the beta function vanishes are shown below.
\begin{center}
$\begin{array}{cc}
N_f&\alpha_c\\
 9 & 5.23599 \\
 10 & 2.20761 \\
11 & 1.2342 \\
 12 & 0.753982 \\
13 & 0.467897 \\
 14 & 0.278017 \\
15 & 0.1428 \\
 16 & 0.0416105 \\
 QED & \alpha \simeq 0.0073 \simeq 1/137
\end{array}$
\end{center}
It is doubtful that we can trust perturbation theory near the perturbative conformal fixed point $\alpha_c$ for $N_f\leq 12$ massless flavors and we need to use the lattice approach.
The cases $N_f$= 8 and 12 have been investigated by lattice practitioners \cite{dgrmp}. 
This is a familiar setup but the asymptotic scaling is much slower than in QCD. 
For $N_f$=12 and $m=0$, some lattice practitioners (e. g. \cite{Hasenfratz:2016dou,Lin:2015zpa}) claim that: 
1) there is no confinement, 2) chiral symmetry is unbroken, 3) the theory is conformal, 
 while others (e. g. \cite{Fodor:2016zil}) claim the opposite. See Ref. \cite{dgrmp} for a critical discussion.

A third logical possibility, confinement  with unbroken chiral symmetry, also called the Wigner mode \cite{benlee}, seems excluded by 't Hooft anomaly matching and decoupling  conditions \cite{tHooft:1979rat}. 
Just the anomaly matching between metaquarks and metabaryons produces very cumbersome solutions. For example for $N_f=8$, the simplest solution we found is (using 
the setup defined precisely  in Ref.  \cite{tHooft:1979rat}):
\begin{center}
$-2\ \begin{Young}
      L&L&L\cr
    \end{Young}-1\  \begin{Young}
      L\cr L\cr L\cr
    \end{Young} 
    -1\ (\begin{Young}
      L\cr \end{Young}\bigotimes \begin{Young}
      R&R\cr
    \end{Young}-\begin{Young}
      R\cr \end{Young}\bigotimes \begin{Young}
      L&L\cr
    \end{Young})$
    \vskip10pt
   $ +2(\begin{Young}
      L\cr \end{Young}\bigotimes \begin{Young}
      R\cr R\cr
    \end{Young}-\begin{Young}
      R\cr \end{Young}\bigotimes \begin{Young}
      L\cr L\cr
    \end{Young})+2\ \begin{Young}
      L&L\cr
      L\cr
  \end{Young}$
  \end{center}
For $N_f=12$, we found no solutions if the 5 indices are less than 12 in absolute value.
If the decoupling condition is added \cite{tHooft:1979rat}, there are no solutions for $N_f>2$. In summary the Wigner mode does not seem a viable alternative. 

Different parts of the RG flows can be connected to hypothetical physical properties. Schematically, the RG flows of asymptotically free gauge theories with a nontrivial IR fixed point have three distinct parts. 
\begin{enumerate}
\item
Setting the scale (initial conditions, far UV)

For a sufficiently high scale we can use the universal perturbative running/dimensional transmutation. 
The QCD analog is $\alpha_s(M_Z^2)\simeq0.1$. This is a physical input. 
\item The intermediate scale

Using the reference scale in 1) and evolving the RG flows, we then reach a physical scale (say in TeV) where we are far from both 
fixed points. 
From a computational point of view, things look maximally nonlinear/multidimensional 
in both directions. 
It is challenging  to capture the essential features with small lattices and 
one-dimensional RG flow approximations. 
\item
The deep IR scale (assuming $m=0$ and an attractive IRFP)

As we continue  the RG evolution, most of the irrelevant features get washed out and 
if we run all the way down to the IRFP, the intermediate scale does not appear anymore. 
\end{enumerate}
For model building applications, the standard model interactions will break the conformal symmetry at the EW scale or a mass can be introduced in order 
to have a confining theory below the explicit chiral symmetry breaking scale.

\section{Progress towards  a linear sigma model description}

Recently, light $\sigma$ masses were found for $SU(3)$ with 
$N_f$=8 \cite{Aoki:2014oha,Appelquist:2016viq,Aoki:2016wnc,Gasbarro:2017fmi} and 12 \cite{Aoki:2013zsa,Aoki:2016wnc} fundamental flavors and for 2 sextets \cite{Fodor:2015vwa}. At the DPF conference, I considered the possibility of using a {\it linear} sigma model to describe the low energy behavior of the massive multiflavor theories. 
It is commonly believed that for the linear model the mass of the $\sigma$ is significantly larger than the mass of the pions for $N_f$=12. This seems in clear contradiction with the the numerical results finding the $\sigma$ lighter than the pions \cite{Aoki:2013zsa,Aoki:2016wnc}. However if we consider models used to describe the 
breaking of the axial $U(1)_A$ in QCD \cite{PhysRevD.3.2874,PhysRevD.21.3388,tHooftphysrep,Meuricea0}, this is not necessarily the case. In a nutshell:
\beq
\phi=(S_j+iP_j)\Gamma_j,
\enq
with $S_0:\sigma$, $P_0:\eta'$ etc., and $Tr(\Gamma_i\Gamma_j)=(1/2)\delta_{ij}$ (for  3 flavors: $j=0,1,2, \dots 8$). The Lagrangian density reads:
\beq
{\mathcal{L}}=Tr\partial_\mu\phi\partial^\mu\phi^\dagger-V(M)-\chi(det\phi + det\phi^\dagger)-bS_0,
\enq
with 
\beq
V( \phi^\dagger \phi)\equiv -\mu^2 \phi^\dagger \phi +Tr \phi^\dagger \phi +(\lambda_1/2-\lambda_2/3)(Tr \phi^\dagger \phi)^2+\lambda_2Tr( (\phi^\dagger \phi)^2). 
\enq
Using the mass formulas of \cite{Meuricea0} for $N_f=3$ in the $SU(3)_V$ limit (3 equal masses), we obtain:
\begin{eqnarray}
\label{eq:main3}
M_{\etp}^2-M_\pi^2&=&(3/2)\chi f_\pi        \nonumber\\
M_\sigma^2-M_\pi^2&=&  \lambda_1f_\pi^2-(1/2)\chi f_\pi\\
M_{a0}^2-M_\pi^2&=&   \lambda_2 f_\pi^2+\chi f_\pi  \nonumber .
\end{eqnarray}
This result shows that the anomaly term provides a negative contribution to the sigma mass. After the conference, this calculation was generalized for arbitrary $N_f$ \cite{lsm}. The mass relations for arbitrary $N_f$ read:
\begin{eqnarray}
\label{eq:main}
M_{\etp}^2-M_\pi^2&=&Xv^{N_f-2}\nonumber\\
M_\sigma^2-M_\pi^2&=&\ls v^2-(1-2/N_f)Xv^{N_f-2}\\
M_{a0}^2-M_\pi^2&=&\la v^2+(2/N_f)Xv^{N_f-2}\nonumber .
\end{eqnarray}
We see that there the anomaly term give a negative contribution to $M_\sigma^2-M_\pi^2$. In addition, it is argued that $\ls v^2/M_{\etp}^2$ varies slowly and that simple formulas such as $M_\sigma^2\simeq (2/N_f-C_\sigma)M_{\etp}^2$ should apply in the chiral limit \cite{lsm}. These results, triggered by the DPF conference, suggest 
lattice calculations that could improve our understanding of the boundary of the conformal window.

\section{End point in the $(\beta,m)$ plane for $N_f=12$}

Numerical calculations for $N_f$=12 have shown that a  line of first order transitions in the $(\beta,m)$ plane has an end point where one expects a 
second order phase transition with a light scalar in its vicinity \cite{Jin:2013hpa}. If this is a lattice artifact or something that could have 
a counterpart in the continuum is an open question. Calculations of the
zeros of the partition function in the complex $\beta=6/g^2$ plane (Fisher zeros) have been performed \cite{zech1,zech2}. 
The scaling of the lowest zeros with $L$ for $N_f=12$ is $Im\beta\propto L^{-4}$ is consistent with a first order transition. 
By increasing the mass we expect to reach the endpoint where a second order transition takes places (4D Ising universality, with a light weakly interacting $0^+$). At the critical mass we would expect $Im\beta\propto L^{-1/\nu}$ with $\nu=1/2$ instead of $L^{-4}$. This question is being investigated by Z. Gelzer and D. Floor. 
Preliminary results \cite{zth} place this endpoint within the range $am_f =[0.05,0.09]$.
The unconventional continuum limit near this endpoint should be studied. 

 \section{Computational methods for near conformal systems}

Investigating approximate conformal symmetry on small lattice is sometimes problematic. 
The Tensor Renormalization Group (TRG) method is an 
ideal tool to study nearly conformal systems. The TRG 
has exact blocking formulas for spin and gauge models \cite{prd88}. It applies to many lattice models: the Ising model, $O(2)$ model, $O(3)$ model, $SU(2)$ principal chiral model (in any dimensions), Abelian and $SU(2)$ gauge theories (1+1 and 2+1dimensions).
One unique feature is that the blocking separates the degrees of  freedom inside the block (integrated over), from those kept to communicate with the neighboring blocks as illustrated in two dimensions below. 
\begin{center} 
     \includegraphics[width=2.in]{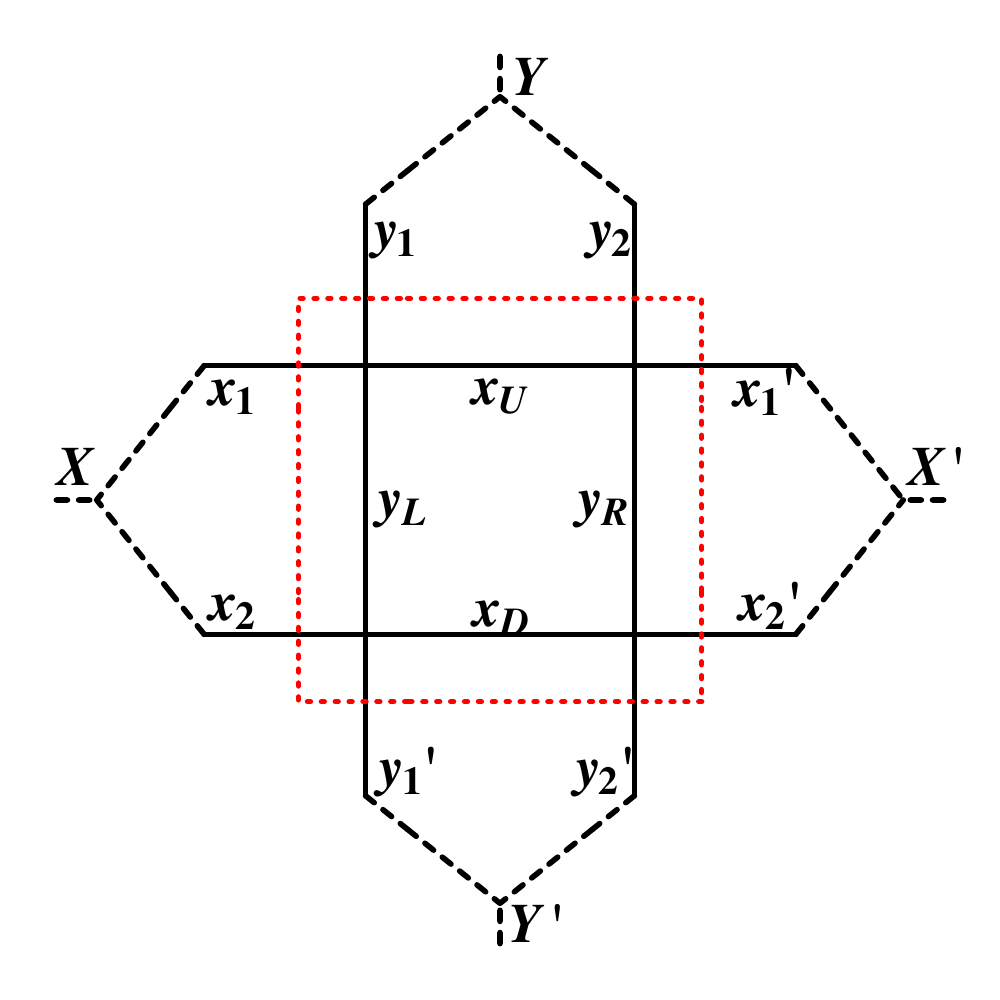}
\end{center}
The only approximation is the truncation in the number of ``states" kept. Approximate fixed points have been found but so far the effect of the truncations on critical exponents is not very well understood \cite{PhysRevB.87.064422,efratirmp}.  The TRG seems free of sign problems and can handle complex temperature and real chemical potential  \cite{prd89,PhysRevA.90.063603}.

The TRG has been used to calculate the entanglement entropy \cite{PhysRevE.93.012138}  in 1+1 dimension and used to check the 
Calabrese-Cardy scaling \cite{Calabrese:2004eu,Calabrese:2005zw} for a spacial volume $N_s$ with open BC:
\begin{equation}
S_n(N_s)=K+\frac{c(n+1)}{12n}\ln(N_s), 
\end{equation}
where $S_n(A)$ is $n$ -th order R\'enyi entanglement entropy of a subsystem $A$ defined as:
\begin{equation*}
S_n(A)\equiv \frac{1}{1-n}\ln(\Tr((\hat{\rho}_A)^n)) \ . 
\end{equation*}

Fits for the central charge for $S_2$ in the $O(2)$ model with a chemical potential are consistent with a central charge $c=1$ \cite{prd96}. 
Quantum simulations with optical lattices have been proposed \cite{pra96} to measure the central charge. 
So far our calculations have been limited to two dimensions (2D). 
Measuring the second-order R\'enyi entropy $S_2$ for 4D gauge theories with fermions seems feasible as it only requires two copies of the subsystem. It could be an interesting approach to the conformal window question. 

\section{Conclusions}
A better understanding of conformal (or near conformal) lattice gauge models is necessary before attempting realistic model building.
Exotic continuum limits (e. g. near the end point of a line of first order transition) should be investigated. 
An effective theory involving the $0^+$ (the $\sigma$) was in progress at the time of the conference  and is now available in a recent preprint \cite{lsm}.
The framework is completely natural and can handle $\Lambda_{new}\gg 2$TeV: experimentalists, keep looking for di-boson resonances! Tensor RG methods are promising to study near conformal spin and gauge models in 2D; they need to be 
extended to 3D and 4D. The entanglement entropy could be an indicator of conformality in gauge theory with fermions.
\section{Acknowledgements:} 
We thank Y. Aoki,  C. Bernard, T. DeGrand, G. Fleming, D. Nogradi, E. Rinaldi, D. Schaich, B. Svetitsky and O. Witzel for useful conversations and suggestions.  
This research was supported in part  by the Department of 
under Award Numbers DOE grant DE-SC0010113. Some of the calculations were done using the NERSC facilities.


  \end{document}